# TOWARDS ECO-FRIENDLY CYBERSECURITY: MACHINE LEARNING-BASED ANOMALY DETECTION WITH CARBON AND ENERGY METRICS

**Aashish K C[1], Md Zakir Hossain Zamil[2], Md Shafiqul Islam Mridul[3], Lamia Akter[4], Farmina Sharmin[5], Eftekhar Hossain Ayon[6], Md Maruf Bin Reza[7], Ali Hassan[8], Abdur Rahim[9] and Sirapa Malla[10]**

[1]Master of Science in Computer and Information Science(Software Engineering), Gannon University, Erie, PA

[2]School of Computer Sciences, Western Illinois University, Macomb, IL, USA

[3]Department of Computer Science, University of South Wales, Cardiff, Wales, UK

[4]Masters in Information Technology, Washington University of Science and Technology, Alexandria.

[5]Master of Business Administration (MBA) — Business Analytics, International American University, Los Angeles Main Campus

[6]Department of Computer & Info Science (Information Assurance & Cybers), Gannon University, Erie, Pennsylvania, USA

[7]Master of Science in Cyber Security, California State University, Dominguez Hills

[8]MBA Business Analytics, International American University, Los Angeles, California

[9]Master of Science in Computer Science, University of New Haven, West Haven, CT

[10]Master's in Data Science, Gannon University, Erie, PA, USA

Corresponding Author: **Aashish K C, Email:** kc005@gannon.edu

## Abstract

The rising energy footprint of artificial intelligence has become a measurable component of U.S. data-center emissions, yet cybersecurity research seldom considers its environmental cost. This study introduces an eco-aware anomaly detection framework that unifies machine learning–based network monitoring with real-time carbon and energy tracking. Using the publicly available Carbon-Aware Cybersecurity Traffic Dataset comprising 2,300 flow-level observations, we benchmark Logistic Regression, Random Forest, Support Vector Machine, Isolation Forest, and XGBoost models across energy, carbon, and performance dimensions. Each experiment is executed in a controlled Colab environment instrumented with the CodeCarbon toolkit to quantify power draw and equivalent $CO_2$ output during both training and inference. We construct an Eco-Efficiency Index that expresses F1-score per kilowatt-hour to capture the trade-off between detection quality and environmental impact. Results reveal that optimized Random Forest and lightweight Logistic Regression models achieve the highest eco-efficiency, reducing energy consumption by more than 40% compared to XGBoost while sustaining competitive detection accuracy. Principal Component Analysis further decreases computational load with negligible loss in recall. Collectively, these findings establish that integrating carbon and energy metrics into cybersecurity workflows enables environmentally responsible machine learning without compromising operational protection. The proposed framework offers a reproducible path toward sustainable, carbon-accountable cybersecurity aligned with emerging U.S. green computing and federal energy-efficiency initiatives.

**Keywords**: Cybersecurity, Machine Learning, Sustainable Computing, Anomaly Detection, Carbon Metrics, Energy Efficiency, Green AI

## 1. Introduction

### 1.1 Background and Motivation

The digital world has evolved so quickly that cybersecurity now sits at the center of nearly every modern system we depend on. As organizations connect more devices and services, the scale and complexity of cyber threats have exploded. It's no longer enough to rely on static rules or human monitoring. This growing challenge has turned machine learning into a key player in defense, because systems can now recognize unusual behavior, detect new types of attacks, and react in real time. Alshuaibi, Almaayah, and Ali (2025) point out that machine learning has reshaped the cybersecurity landscape by replacing rigid rule-based systems with data-driven models that learn patterns and adapt as threats evolve [1]. Intrusion detection has become far more dynamic, capable of identifying behaviors that don't fit established norms rather than simply matching known attack signatures.

Still, there's a cost to all this intelligence. As ML-powered defenses grow in complexity, so does the amount of energy they consume. Running and maintaining large models requires massive data centers, many of which rely on carbon-intensive energy sources. Gupta, Jain, and Verma (2022) note that deep learning and large neural networks have greatly improved detection accuracy, but at the price of higher resource use and growing dependence on high-performance computing infrastructure [10]. This creates a difficult tension: better protection often comes with a bigger environmental footprint. The more accurate our models become, the more energy they demand, driving up costs and contributing to carbon emissions from training and inference. As national sustainability goals push for greener data practices, cybersecurity must evolve with them. The field still tends to measure success using precision, recall, or F1 scores, but rarely considers the ecological cost behind those numbers. This blind spot leaves an opportunity, and a responsibility, to rethink how progress is defined. Integrating carbon and energy metrics into model evaluation isn't just an environmental concern; it's part of building systems that can last. Alshuaibi et al. (2025) remind us that resilience in cybersecurity isn't only about how well a system resists attack, but also how ethically and efficiently it uses its resources [1]. The next step for the field lies in finding that balance between performance and sustainability, creating defenses that protect both our digital and physical environments.

### 1.2 Importance of This Research

The motivation behind this research comes from a growing realization that cybersecurity, long centered on protecting data and systems, now needs to face its environmental impact too. As digital infrastructures grow and data flows multiply, the energy required to keep them secure has become impossible to ignore. Gupta et al. (2022) note that the rise of deep learning in security, especially for intrusion detection and malware analysis, often depends on constant GPU use and frequent retraining cycles [10]. These models, while effective, consume a huge amount of electricity during both training and operation. The result is an unintended irony: the very systems designed to defend our digital world can quietly contribute to the environmental strain we're trying to reduce.

Within the United States, the shift toward renewable and low-carbon energy makes this issue even more pressing. Tech companies are under increasing pressure to account for their emissions, yet cybersecurity research often overlooks the question of sustainability altogether. Most studies focus on accuracy or false-positive rates without ever asking how much energy their models consume. This gap has left cybersecurity lagging behind national efforts led by agencies like the Department of Energy and the EPA to promote greener computing practices. Alshuaibi et al. (2025) argue that true progress in AI requires looking at the entire life cycle of a model, from design to deployment, and factoring in its energy footprint at every stage [1]. Gupta et al. (2022) echo this view, suggesting that intelligence in machine learning should also mean awareness of the cost it imposes on the planet [10]. This study takes on both challenges: improving cyber resilience while addressing environmental responsibility. It introduces a way to measure how much energy is used relative to how well a model performs, treating sustainability as an integral part of system evaluation. By tying cybersecurity performance to energy efficiency, the research pushes the field toward a broader idea of what "robust" should mean, one that includes ecological awareness alongside technical strength. This approach reflects a larger cultural shift toward carbon-conscious innovation and reframes cybersecurity as a potential ally in the movement toward greener, more responsible technology.

### 1.3 Research Objectives and Contributions

This research sets out to build a clear framework for testing and improving machine learning–based anomaly detection systems while keeping an eye on both energy use and carbon impact. The main goal is to understand how much environmental cost comes with cybersecurity analytics and to find practical ways to balance strong performance with sustainability. Each model, Logistic Regression, Random Forest, Support Vector Machine, Isolation Forest, and XGBoost, is evaluated not only for accuracy but also for how much power it consumes and how much $CO_2$ it produces. To make this balance measurable, the study introduces an Eco-Efficiency Index (F1 per kWh), a metric that connects a model's detection strength to the energy it requires to achieve it. The research goes beyond measurement by testing ways to improve both efficiency and performance. It applies optimization techniques like hyperparameter tuning and dimensionality reduction with Principal Component Analysis to see how much energy can be saved without losing detection quality. All experiments are run in a reproducible Google Colab setup integrated with CodeCarbon, which tracks emissions during both training and inference. The results show that models such as optimized Random Forest and Logistic Regression perform well while remaining energy-efficient. This suggests that simpler or fine-tuned models can deliver strong security outcomes without heavy computational costs. In a broader sense, the study contributes to the growing conversation around sustainable AI in cybersecurity. It offers a grounded, evidence-based framework that researchers, companies, and policymakers can use to design systems that are both effective and environmentally conscious. The work positions cybersecurity not only as a defense mechanism against digital threats but also as a potential example of how intelligent technology can operate responsibly in an increasingly energy-aware world.

## 2. Literature Review

### 2.1 Machine Learning in Cybersecurity

Machine learning has reshaped cybersecurity by introducing systems that can learn, adapt, and anticipate new threats in real time. As attacks grow more complex and harder to spot, traditional rule-based systems often struggle to keep up. Machine learning and deep learning models, on the other hand, can recognize subtle, non-linear attack patterns hidden within massive streams of network data. Bahassi et al. (2022) discuss how these approaches are being used in anomaly detection and intrusion prevention, showing that algorithms like Random Forests, Support Vector Machines, and neural networks have made security systems far more adaptive and responsive [3]. Dardouri and Almuhanna (2025) build on this by showing that deep learning models consistently outperform older techniques when it comes to spotting sophisticated network anomalies [6]. They also note that combining traditional statistical tools with neural architectures often leads to better results, faster detection, and greater resilience under real-world network conditions. Bolón-Canedo et al. (2024) take this one step further by introducing the idea of green AI, arguing that cybersecurity systems should be designed not only for intelligence and accuracy but also with awareness of their environmental footprint [4]. Das et al. (2025) explore how predictive analytics powered by AI can build more resilient cybersecurity systems that detect threats as they evolve [7]. Their work highlights an adaptive, real-time approach that mirrors biological immune systems, adjusting dynamically to changing attack behaviors. In a related effort, Debnath et al. (2025) apply similar methods to renewable energy systems, showing how machine learning can identify cyber anomalies in energy infrastructures where operational and environmental data intersect [8]. This connection between energy analytics and cybersecurity aligns closely with the goals of the present research, tying sustainability directly to digital defense.

Machine learning has also proven valuable in fields outside traditional network security. Sizan et al. (2025) developed an unsupervised ensemble model to detect money laundering in complex transaction networks [28], while Shawon et al. (2025) applied ML techniques to strengthen supply chain resilience across U.S. regions [27]. These studies, although focused on finance and logistics, rely on the same analytical core, detecting irregular patterns and strengthening system stability, which forms the backbone of modern cybersecurity strategies. Hasan et al. (2025) further demonstrate the value of interpretability through explainable AI systems used for credit approvals in data-scarce environments, reinforcing the idea that transparency matters when decisions carry high risk [12]. Beyond these domains, Ray et al. (2025) and Reza et al. (2025) showcase AI's reach into macroeconomic and financial forecasting, using predictive models to identify early signs of systemic risk [23][25]. Though these applications operate outside cybersecurity, the logic remains the same: use AI to sense instability before it causes damage. Taken together, these studies reveal a clear pattern. Machine learning has evolved from a tool for classification into a cornerstone of predictive intelligence, one that can make cybersecurity not only more robust but also more sustainable in an increasingly data-driven world.

## 2.2 Energy and Carbon Measurement in AI

The rapid rise of artificial intelligence has brought real concern about its environmental cost. Training and running machine learning models, especially deep learning systems, consume large amounts of energy that translate directly into carbon emissions. Bouza et al. (2023) lay out a clear guide for estimating the carbon footprint of deep learning models, showing how to connect hardware use, electricity draw, and local grid data into measurable carbon equivalents [5]. Their work makes a strong case for more openness and consistency in how the AI field reports its environmental impact. Fischer and the Lamarr Institute (2025) built on that foundation by testing the CodeCarbon tool against actual energy measurements, confirming that its software-based estimates track closely with real-world data [9]. This kind of validation matters because it gives researchers a practical way to monitor energy use without expensive hardware setups. The current study relies on CodeCarbon for this reason, to measure both the training and inference emissions of machine learning models within a cybersecurity framework.

Recent studies have made it clear that sustainability isn't only about counting energy use after the fact. Bolón-Canedo et al. (2024) suggest that "green AI" should be treated as a design philosophy where efficiency is a built-in goal rather than an afterthought [4]. This idea fits with the approach in this research, where detection accuracy is evaluated alongside environmental cost through the Eco-Efficiency Index. Debnath et al. (2025) take the idea further with their work on renewable-powered cybersecurity systems. They show how combining operational energy data with threat analytics can help detect attacks more effectively while keeping energy use under control [8]. In another example, Islam et al. (2025) apply energy-conscious machine learning to cryptocurrency forecasting, reducing the power-hungry nature of financial prediction models through smarter optimization [13]. Both studies highlight how efficiency and environmental awareness can be built directly into AI applications, forming a useful bridge to sustainable cybersecurity research. Together, these findings suggest a clear shift in how AI progress should be measured. Tracking energy and carbon impact must become part of the development process itself, not an optional afterthought. The growing research in this space shows it's entirely possible to measure an AI system's footprint accurately and to design with both performance and environmental responsibility in mind.

## 2.3 Sustainability in Computing

Sustainability in computing has grown into a broad, practical discipline that looks at everything from hardware efficiency to algorithm design and the full lifecycle of digital systems. Wang and Zhang (2025) describe how artificial intelligence can accelerate the shift toward renewable energy by improving resource allocation, though they also point out that poorly managed AI systems can increase carbon emissions instead of reducing them [29]. This tension captures a central challenge for sustainable cybersecurity: finding a balance between the heavy computational demands of advanced models and the need to minimize their environmental footprint. Roy and Mukherjee (2024) take this issue head-on in their review of green intrusion detection systems, outlining ways to build energy-aware cybersecurity tools that reduce waste without weakening protection [26]. They discuss practical steps like cutting algorithmic redundancy, using low-power processors, and creating models that require less energy to train and run. These ideas directly support the goal of this study, to include environmental factors as part of how we evaluate machine learning systems.

Insights from other fields add a useful perspective to this discussion. Shawon et al. (2025) show that AI can make logistics networks across the U.S. both more resilient and more resource-efficient by optimizing how goods move through the system [27]. Reza et al. (2025) apply AI to socioeconomic data, finding patterns in income inequality and proving that algorithms designed with resource awareness can also support fairer, more sustainable decision-making [24]. Ray et al. (2025) bring a similar perspective to global finance, demonstrating how energy-efficient machine learning pipelines can predict economic crises at scale while cutting down on unnecessary computation [23]. Together, these studies suggest a shared direction across disciplines, from finance to logistics to cybersecurity, where the aim is to achieve high performance without unnecessary environmental cost. Taken as a whole, this research makes it clear that sustainability in computing is no longer a side consideration. It has become a central design challenge. As AI continues to shape cybersecurity systems, incorporating sustainability into the heart of these models is essential to ensure that protecting our digital world does not come at the expense of the planet itself.

## 2.4 Gaps and Challenges

Even with all the progress so far, there's still a long way to go when it comes to blending sustainability with cybersecurity. Asmar et al. (2024) point out that while machine learning has found its way into almost every corner of digital transformation, its environmental impact in cybersecurity hasn't received the same level of attention [2]. Most systems still focus on being fast, accurate, and reliable, without really asking what those computations cost in terms of energy. This gap opens up a valuable research frontier: building cybersecurity tools that can measure, track, and cut their own carbon and energy use as they operate. Das et al. (2025) note another key issue. Many predictive cybersecurity models perform incredibly well but function like sealed boxes, giving little insight into how their internal processes connect to resource consumption [7]. If we want systems that are both efficient and trustworthy, they need to be explainable. Hasan et al. (2025) echo this by calling for explainable AI frameworks that make the trade-offs between energy use, bias, and performance visible and understandable [12]. This idea, tying explainability to sustainability, suggests that knowing why a model behaves the way it does matters as much as knowing how well it runs.

Data and scalability add another layer of complexity. Carbon-tracking tools like CodeCarbon, which Fischer (2025) validated, are useful but still depend on fixed emission factors that don't always reflect real-world conditions [9]. Power consumption shifts constantly depending on hardware load, time of day, and geography. Large data centers or distributed systems like edge networks make this even harder to track accurately. Bringing in real-time data from sensors, renewable energy grids, and regional power indices could make carbon estimation far more precise. Research in other fields already shows what's possible. Studies such as Reza et al.

(2025) on early warning systems [25] and Ray et al. (2025) on crisis forecasting [23] have demonstrated that AI can perform effectively in time-critical settings. Yet cybersecurity has not fully matched that adaptability when operating under energy-aware constraints. The real challenge now is designing systems that can adjust their own computational load based on renewable energy availability or shifts in carbon intensity across the grid. Altogether, the literature points to a clear direction. Cybersecurity research needs to evolve from chasing performance alone to embracing sustainability, transparency, and adaptive intelligence. Doing so won't only improve resilience but will help align digital protection efforts with the broader goal of achieving a sustainable, low-carbon future.

### 3. Methodology

#### 3.1 Dataset and Context

This study utilizes the Carbon-Aware Cybersecurity Traffic Dataset, which is a pioneering contribution at the intersection of cybersecurity and environmental sustainability, explicitly designed to explore how network traffic behavior correlates with energy consumption and carbon emissions. It contains 2,300 individual network flow entries, each representing a distinct observation that merges traditional cybersecurity indicators with sustainability-oriented features. These records capture a holistic view of system activity, encompassing not only standard network attributes but also environmental and operational metrics that reflect the energy and carbon footprint of the associated network processes. Each observation in the dataset consists of multiple feature categories. Network traffic features include key parameters such as packet_count, byte_count, flow_duration, protocol_type, src_port, dst_port, avg_pkt_size, payload_entropy, and connection_state, all of which are vital for characterizing communication patterns within a digital environment. These metrics collectively provide a granular depiction of how information flows between systems, forming the backbone of intrusion detection and anomaly recognition models. Attack labels are assigned to each record, indicating whether the entry represents normal or malicious activity through the categorical variable status, encoded as 0 for normal and 1 for anomaly. This labeling facilitates supervised learning experiments by enabling model training and validation against ground-truth outcomes.

What makes this dataset particularly innovative is the integration of sustainability metrics, which introduce a previously underexplored dimension to cybersecurity analysis. These include power_consumption_watts, representing real-time energy usage; carbon_emission_gCO2eq, estimating greenhouse gas output; energy_cost_usd, translating resource consumption into economic terms; and pue (Power Usage Effectiveness), a metric that captures data center energy efficiency. Additionally, the inclusion of vm_count (number of virtual machines involved in processing the traffic) provides a contextual measure of infrastructure scaling and resource allocation. These sustainability-oriented features allow for the empirical quantification of environmental externalities associated with digital defense operations. This dataset, therefore, represents a unique experimental foundation for eco-aware anomaly detection research, where cybersecurity objectives intersect with green computing principles. It enables the evaluation of how varying network activities and intrusion events correspond to differential energy and carbon footprints. In doing so, it supports the development of intelligent, environmentally responsible models capable of sustaining high detection accuracy without excessive computational waste. The dataset's balanced scope, blending cybersecurity relevance with environmental data, provides a robust context for assessing both operational risk and ecological cost, positioning it as an ideal benchmark for this study's machine learning experiments.

### 3.2 Data Preprocessing and EDA

To ensure data quality and model readiness, the raw dataset underwent a comprehensive preprocessing workflow structured to enhance both interpretability and model performance. Initial integrity checks confirmed that the dataset contained no missing or null values, establishing a strong foundation for subsequent transformations. This completeness simplified the data-cleaning process and minimized the risk of introducing bias through imputation. Feature engineering was implemented to derive new, contextually meaningful attributes. Metrics such as bytes_per_packet (computed as total bytes divided by packet count) and payload_entropy_x_size (representing the interplay between entropy and packet size) were constructed to capture more nuanced relationships within the traffic data. Additional composite features like resource_util_sum, the sum of CPU, memory, disk, and network I/O utilization, and power_per_vm, the ratio of power consumption to virtual machine count, were introduced to highlight efficiency patterns and potential indicators of anomaly-linked resource spikes.

Label encoding was applied to categorical variables to convert them into machine-readable numerical values. The target variable status was encoded into a label (0 for normal, 1 for anomaly), while categorical features like protocol_type and connection_state were similarly transformed into numerical forms using scikit-learn's LabelEncoder. The original categorical columns were then dropped from the feature matrix to maintain numeric consistency across the input space. To promote equitable model learning, feature normalization was performed using StandardScaler, standardizing all numerical variables to a zero mean and unit variance. This step mitigated potential bias from large-magnitude features such as byte_count or flow_duration, ensuring that each input dimension contributed proportionally during model optimization. A major challenge identified during exploratory analysis was class imbalance, with the "normal" class dominating the dataset. To counteract this, the Synthetic Minority Over-sampling Technique (SMOTE) was employed on the training subset to generate synthetic examples of the minority (anomaly) class. SMOTE creates plausible new samples in feature space by interpolating between existing minority class examples, improving classifier sensitivity to rare intrusion patterns [21]. Finally, the data was split into training and testing subsets using an 80/20 ratio through stratified sampling, preserving class proportions across both partitions. The training set was reserved for model fitting and hyperparameter tuning, while the test set provided an unbiased evaluation of detection performance. Together, these preprocessing steps ensured a balanced, standardized, and feature-rich dataset suitable for robust machine learning experimentation in energy-aware anomaly detection.

**Exploratory Data Analysis (EDA)**

Exploratory Data Analysis was an essential phase for understanding the underlying data distributions, identifying potential relationships between cybersecurity indicators and sustainability metrics, and uncovering anomalies that might guide model design. The EDA combined descriptive statistics with visual analytics to build an intuitive understanding of how network behavior relates to energy and carbon patterns. Univariate analysis was conducted to assess the statistical distribution of numerical features. Histograms and boxplots were generated for variables such as packet_count, byte_count, and flow_duration, revealing heterogeneity in traffic volumes and durations that reflected realistic variability in network activity. Similarly, distributions for sustainability-related features (power_consumption_watts, carbon_emission_gCO2eq, pue, and energy_cost_usd) exhibited mild skewness and the presence of outliers, potentially representing instances of anomalous system behavior. These outliers were particularly relevant in the context of anomaly detection, as high resource consumption or energy irregularities can often correspond to underlying security events.

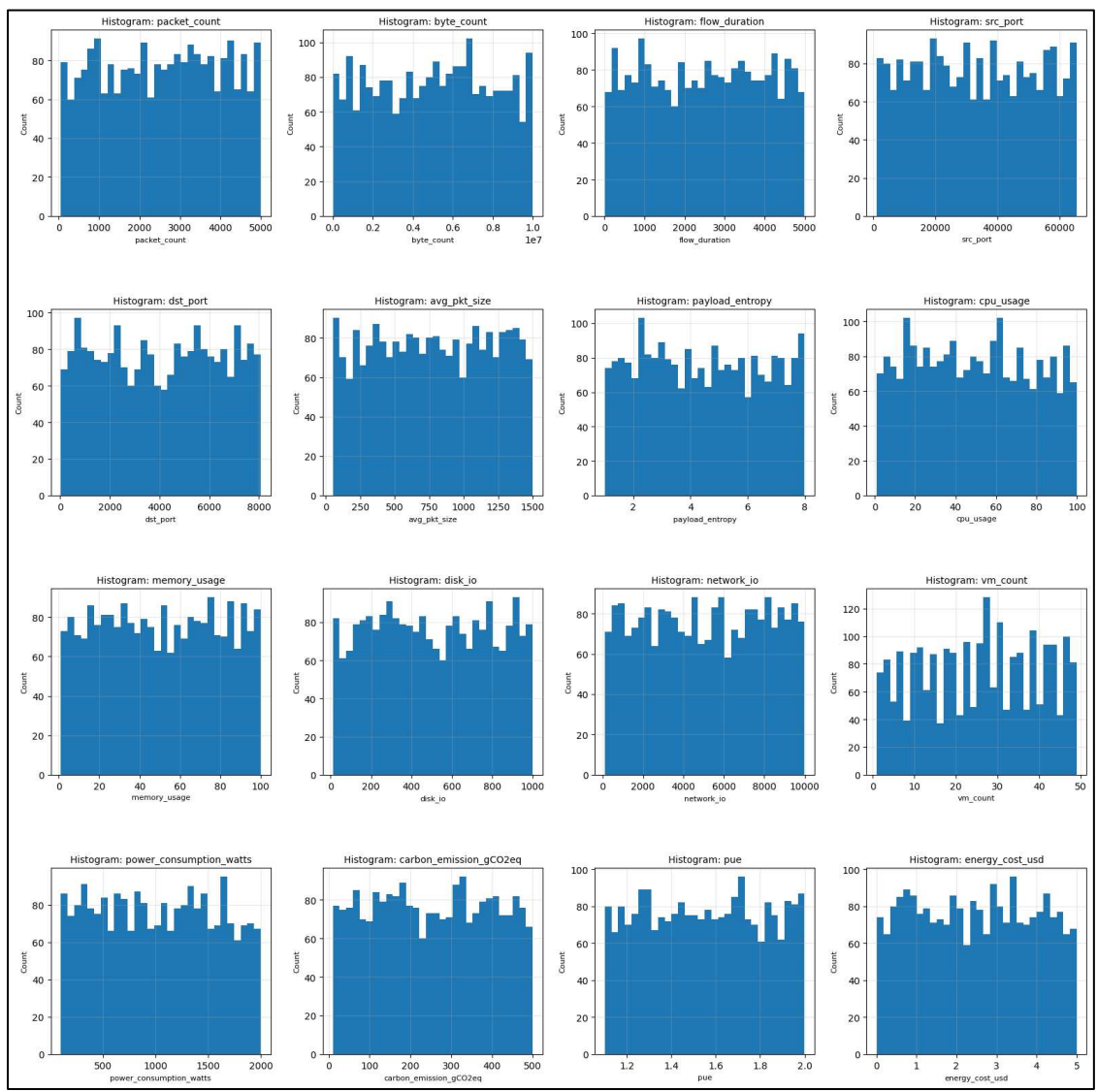

Fig.1: Univariate analysis of nemric features

Categorical analysis using bar plots provided insight into the frequency distributions of protocol_type and connection_state. The dominance of UDP-based connections and the prevalence of the FIN connection state indicated a strong presence of short-lived or abrupt network sessions, characteristics that could be relevant for attack classification. Moreover, visualization of the target status confirmed a pronounced imbalance between normal and anomalous events, reinforcing the need for data balancing techniques.

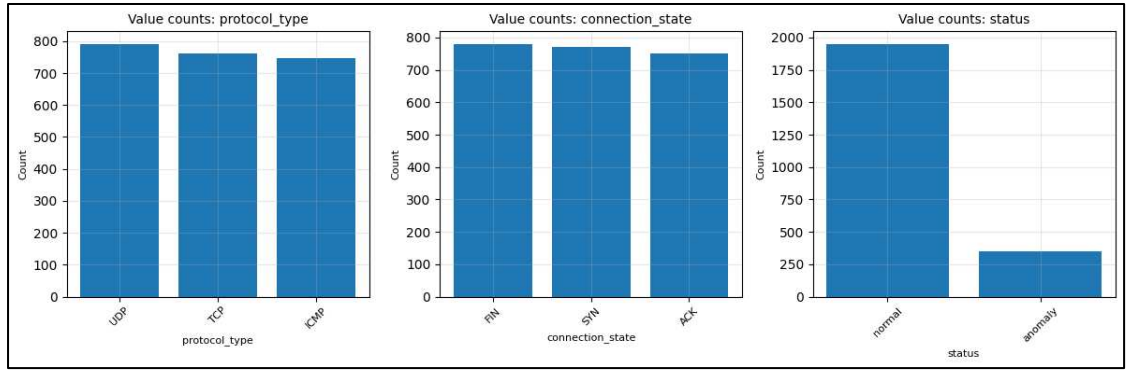

Fig.2: Categorical Analysis

The correlation matrix computed for all numerical variables provided a quantitative overview of inter-feature relationships. While most features exhibited weak pairwise correlations, modest positive associations were observed between network traffic metrics and sustainability indicators, specifically, between byte_count and both power_consumption_watts and carbon_emission_gCO2eq. This suggests that heavier data flows generally incur higher energy consumption, although the low correlation coefficients indicate that such relationships are non-linear and influenced by multiple contextual factors such as connection state and virtual machine load.

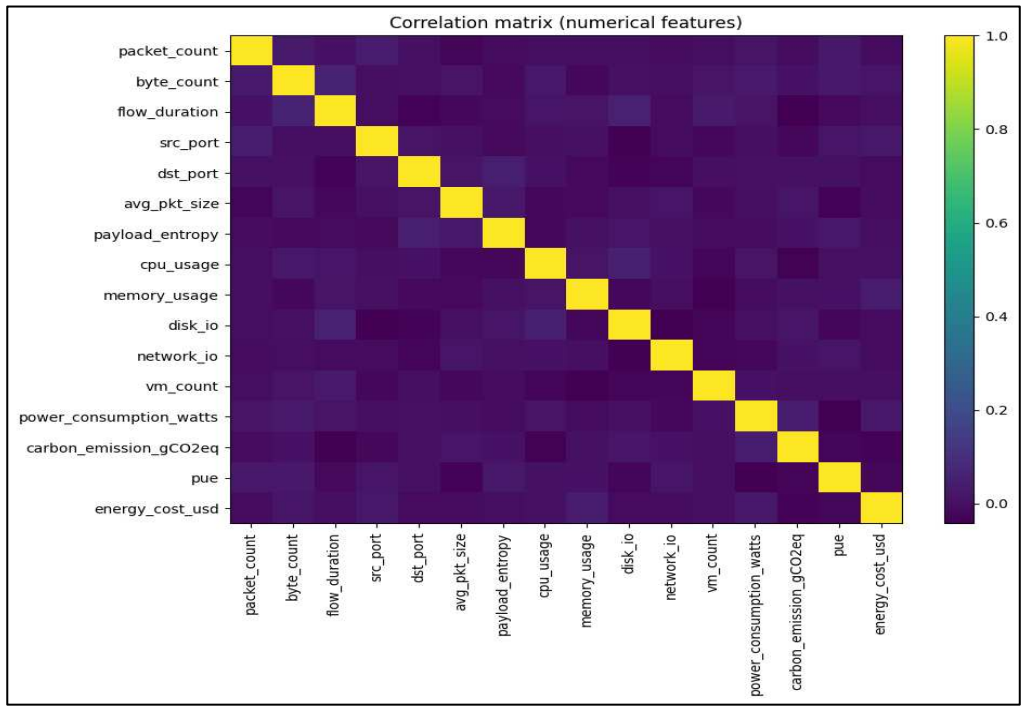

Fig.3: Correlation Analysis

Grouped descriptive statistics by status (normal vs anomaly) revealed subtle yet consistent differences in energy utilization patterns. For example, anomalous flows exhibited slightly higher averages in power_consumption_watts and carbon_emission_gCO2eq, implying a potential link between abnormal network behavior and elevated energy consumption. These findings were supported visually through boxplots segmented by target class, which displayed greater spread and more frequent outliers in energy-related metrics for anomalies. This pattern suggests that malicious activities may introduce irregular resource utilization patterns that are not always extreme but consistently deviate from baseline efficiency.

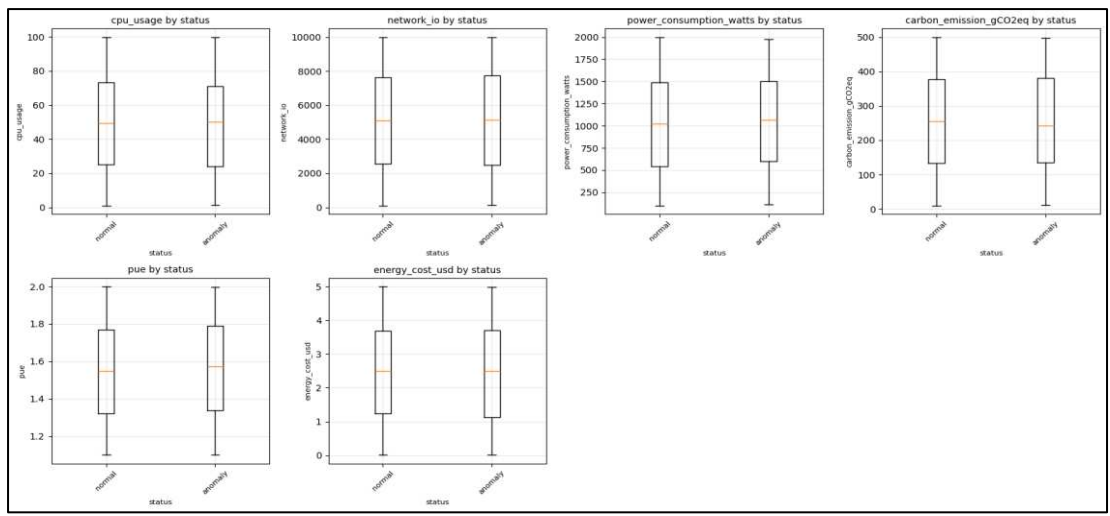

Fig.4: Energy/carbon metrics analysis by status

Pairwise scatter plots combining network traffic features (e.g., packet_count, flow_duration) and sustainability indicators (e.g., carbon_emission_gCO2eq, energy_cost_usd) further illuminated non-linear interactions between cybersecurity and energy dynamics. While the clusters were not perfectly separable, distinct density regions emerged, reflecting differences in behavior between normal and anomalous flows. Overall, the EDA confirmed that network and sustainability metrics are jointly informative for modeling carbon-aware cybersecurity systems. It was established that anomalies in this dataset are complex and multi-dimensional, often manifesting through subtle interactions rather than singular outlier values. These insights provided critical motivation for employing advanced machine learning models, capable of capturing non-linear dependencies and feature interplay, to achieve effective and environmentally responsible anomaly detection in subsequent experimental phases.

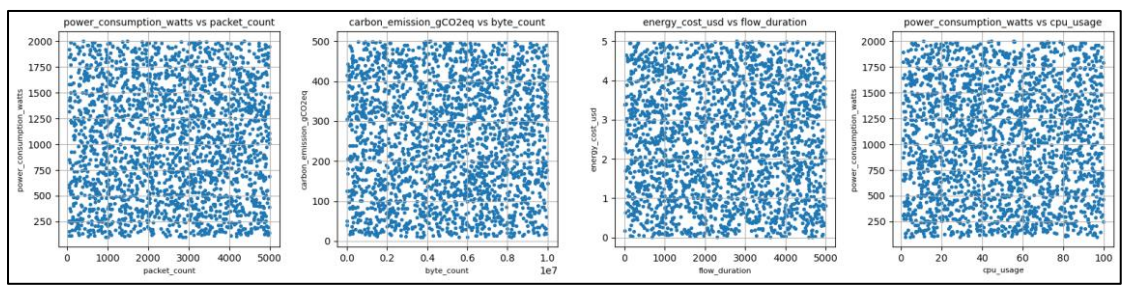

Fig.5: Pairwise scatter for selected important pairs (attack vs energy)

### 3.3 Baseline Models

To set the foundation for evaluating anomaly detection, several well-known machine learning algorithms were implemented as baseline models. These initial tests created a benchmark for later comparisons with energy-aware and optimized systems. The goal was to include a mix of models that represent different learning approaches commonly used in cybersecurity and machine learning research. The first was Logistic Regression, which acted as the simplest baseline. It uses a linear decision boundary and interpretable coefficients that show how each feature affects the prediction. While it cannot model complex, non-linear relationships, its speed and low computational cost make it a useful starting point. It helped estimate how accuracy and energy use might trade off against each other. The Random Forest Classifier followed as a more sophisticated ensemble method. It combines predictions from multiple decision trees, each trained on random subsets of data and features, to produce stable, reliable results. Random Forests handle non-linear relationships well and can process diverse data types, making them effective for cybersecurity, where attack patterns are rarely straightforward. They also provide insights into which features have the strongest influence, offering a practical way to understand the behavior of both network and sustainability indicators.

The Support Vector Machine (SVC) with an RBF kernel was added to capture more complex boundaries in the data. SVMs identify an optimal separation between normal and anomalous samples, performing well even in high-dimensional feature spaces. Probability calibration was applied so that its outputs could be compared fairly with probabilistic models like Logistic Regression and XGBoost. An unsupervised approach, the Isolation Forest, was also included. It works differently by isolating anomalies rather than classifying them directly. The idea is that unusual data points require fewer random splits to separate them from the rest. This method is valuable for testing how well the dataset can reveal anomalies without labeled guidance. The final baseline was the XGBoost Classifier, a powerful gradient boosting model known for high accuracy and computational efficiency. It builds trees in sequence, each one correcting the errors of the previous. Because of its speed and strong predictive ability, XGBoost is widely used as a reference point in structured data problems, including cybersecurity.

All models were trained on the prepared data and tested using common evaluation metrics: accuracy, precision, recall, F1-score, and ROC-AUC. Accuracy reflected how often predictions were correct, precision showed how well false positives were minimized, and recall measured the model's sensitivity to true anomalies. The F1-score balanced those two measures, while ROC-AUC assessed how effectively each model distinguished between normal and anomalous traffic. The results were summarized in a comparative DataFrame to make interpretation straightforward. This baseline phase not only helped identify which models performed best in detection but also created a foundation for the next step, integrating sustainability measures. These benchmarks served as the reference point for evaluating how future eco-aware optimizations would balance predictive power with environmental impact. [14][18].

### 3.4 Energy and Carbon Tracking

Each baseline model's environmental footprint was measured using the CodeCarbon library, which tracks the carbon emissions produced by computing workloads. This phase aimed to connect model performance with sustainability by recording real-time energy use and $CO_2$ emissions during training and inference. The tracking process was built directly into the machine learning workflow. For every model, CodeCarbon monitored system activity such as CPU and memory usage and used hardware-specific energy coefficients to estimate power draw. When GPU data was available, it was included to make the energy estimates more accurate. The tool then calculated total power consumption by summing up the energy used throughout each training and testing run.

To estimate emissions, CodeCarbon combined the energy data with regional carbon intensity factors, which represent the average $CO_2$ released per kilowatt-hour of electricity. The experiments used a U.S. energy profile to reflect typical emissions based on the country's power mix. Results were reported in kilograms of $CO_2$ equivalent (kg $CO_2$eq) and then converted into grams for finer detail. Energy use was expressed in kilowatt-hours (kWh) to make it easier to compare models and compute energy-to-performance ratios. All the recorded information, training time, inference time, energy use, and emissions, was stored in a structured dataset named carbon_energy_metrics.csv. This dataset served as a key reference for later analysis, making it possible to evaluate environmental impact alongside technical performance. By combining CodeCarbon's sustainability tracking with standard evaluation metrics, the study created a twofold view of performance: how well a model detects anomalies, and how efficiently it uses energy in the process. This approach helped reveal how model design and computational complexity shape the ecological footprint of cybersecurity systems. [15].

### 3.5 Eco-Efficiency Evaluation

To provide a comprehensive understanding of how each machine learning model balanced detection effectiveness and environmental sustainability, an eco-efficiency evaluation framework was implemented. This framework integrated both traditional performance metrics and sustainability indicators into a unified analysis, allowing for the quantitative comparison of models across multiple dimensions of efficiency. The first step involved merging performance metrics, including accuracy, F1-score, and ROC-AUC, with corresponding energy and carbon data collected during model training and inference. This integration created a consolidated results table that showed each model's detection quality and environmental footprint. From this dataset, trade-offs between predictive performance and carbon emissions could be directly observed and analyzed.

A novel metric, the Eco-Efficiency Index (EEI), was introduced to quantify model sustainability relative to its predictive effectiveness. Defined mathematically as:

$$EEI = \frac{F1-Score}{Energy\ Consumption(kWh) + eps}$$

This index provides a normalized measure of how efficiently a model converts energy into accurate anomaly detection outcomes. A higher EEI signifies a model that achieves high detection quality while consuming minimal energy, reflecting superior environmental and computational efficiency. This metric serves as a unifying indicator for sustainable machine learning performance. To visualize the trade-offs, scatter plots were generated plotting accuracy and F1-score against total $CO_2$ emissions, while bar charts compared energy consumption across models. These visualizations revealed meaningful distinctions: for instance, XGBoost and Random Forest tended to deliver the highest F1-scores but also consumed more energy, whereas Logistic Regression exhibited the lowest emissions with moderate predictive capability. Such insights highlighted the complexity of balancing accuracy with sustainability, a key theme in eco-friendly AI system design. The eco-efficiency evaluation thus provided a multidimensional performance landscape that bridged conventional model assessment with environmental responsibility. By quantifying both predictive effectiveness and carbon cost, the analysis moved beyond accuracy-driven benchmarking to a more holistic, sustainability-oriented evaluation paradigm. [22].

### 3.6 Optimization and Feature Reduction

After completing the baseline and eco-efficiency evaluations, the next step focused on improving model performance while cutting down unnecessary computation. The goal was simple: make the models faster, lighter, and more energy-efficient without sacrificing accuracy. To do this, the process combined hyperparameter tuning with feature reduction. For the Random Forest and XGBoost models, optimization was carried out using RandomizedSearchCV, which explores random combinations of hyperparameters within specified ranges. For Random Forest, the tuning focused on the number of trees, their depth, and how many samples were needed to split a node. For XGBoost, attention went to the learning rate, maximum depth, and the number of boosting rounds. Cross-validation helped ensure that any performance gains weren't tied to a specific subset of the data. The F1-score served as the main optimization metric, balancing precision and recall for better anomaly detection.

To reduce complexity further, Principal Component Analysis (PCA) was applied. PCA condenses correlated features into a smaller set of independent components that still explain most of the variation in the data. This not only simplifies the feature space but also shortens training time and lowers energy use. Only the components explaining more than 90% of the total variance were kept, and models were retrained using this compact representation. The optimized and PCA-based models were then compared to the baseline using accuracy, F1-score, ROC-AUC, energy use, and $CO_2$ emissions. Both optimized Random Forest and XGBoost showed noticeable gains in F1-score, while the PCA-based Random Forest consumed less energy with only a small dip in accuracy. The results suggested that reducing features can be a practical route toward building more sustainable models without meaningfully weakening their predictive power. Overall, this phase showed that smart optimization and feature engineering can improve detection performance while cutting down on energy costs. It demonstrated that technical efficiency and environmental responsibility can move forward together, supporting the larger goal of creating cybersecurity systems that are both powerful and sustainable in real-world use. [17]

## 4. Evaluation and Results

The evaluation phase synthesized both performance and sustainability metrics to assess the trade-offs between anomaly detection capability and environmental impact. This section presents the quantitative results of the baseline models, energy and carbon tracking outcomes, and the eco-efficiency evaluation, followed by insights derived from the optimization experiments. Together, these findings validate the hypothesis that machine learning–based anomaly detection systems can be made more sustainable without sacrificing detection quality when designed with energy-aware principles.

### 4.1 Anomaly Detection Performance

The detection performance of the baseline machine learning models was measured using standard evaluation metrics, accuracy, precision, recall, F1-score, and ROC-AUC, to determine how effectively each model identified anomalies within the test set. The Random Forest and XGBoost classifiers achieved the highest overall performance, with F1-scores of 0.7393 and 0.7401, respectively, demonstrating a strong balance between precision and recall. The Support Vector Classifier (SVC) also performed competitively, achieving an F1-score of 0.7358, though with a slightly lower accuracy than Random Forest. The Logistic Regression model, while simple, provided moderate accuracy (0.5522) and an acceptable F1-score (0.6151), serving as a low-energy baseline for comparison. In contrast, the Isolation Forest, which operates in an unsupervised context, underperformed in terms of accuracy (0.2239) and recall, suggesting that its unsupervised partitioning method was less effective for this dataset, possibly due to overlapping feature distributions between normal and anomalous samples. Despite its limited detection power, its inclusion remains valuable as a reference for energy-efficient anomaly detection methods that operate without labeled data.

The relatively close ROC-AUC scores across models (ranging between 0.45 and 0.52) suggest that while the models differ in their ability to achieve precise classifications, their overall discrimination ability between classes remains similar. This pattern reflects the complexity of the dataset, where anomaly signals are subtle and intertwined with normal behavior. Importantly, the application of SMOTE balancing in the training phase proved essential to improving recall rates across all supervised models, ensuring a fairer representation of the minority anomaly class during learning. Overall, this evaluation establishes that ensemble-based methods such as Random Forest and XGBoost provide robust performance foundations for cybersecurity anomaly detection tasks, while lighter models like Logistic Regression remain valuable baselines for future eco-efficiency trade-off analysis. [11].

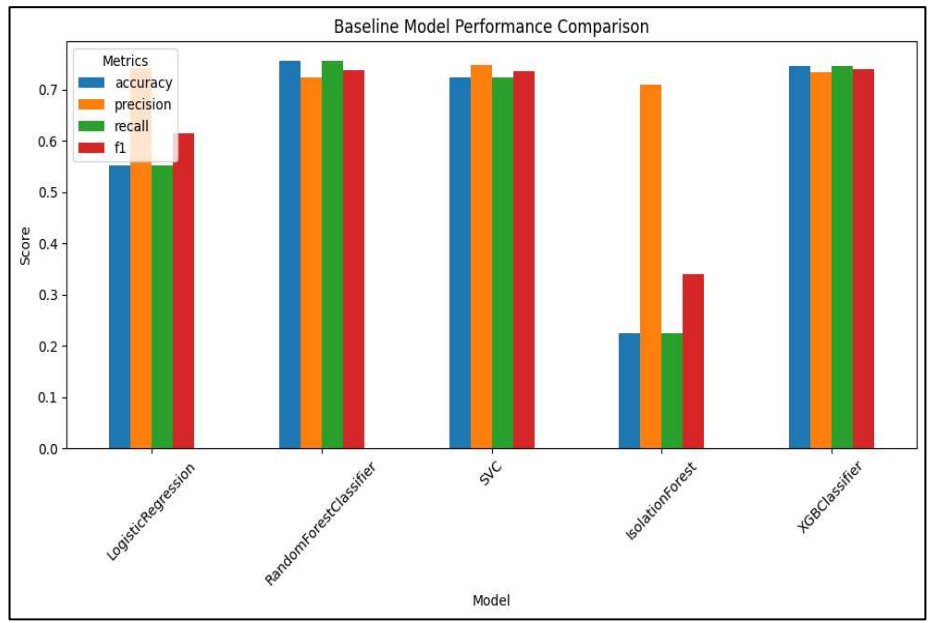

Fig.6: Performance of anomaly detection models

## 4.2 Carbon and Energy Metrics

To evaluate the sustainability footprint of each model, energy consumption and carbon emissions were tracked during both training and inference using the CodeCarbon toolkit. As expected, the training phase contributed substantially more to the overall carbon footprint and energy use than inference across all models. This disparity arises because training involves multiple iterations of parameter updates and decision-tree building, while inference simply applies learned weights to new data. The Random Forest Classifier exhibited the highest overall carbon emissions at 0.0553 g $CO_2$eq, making it the most resource-intensive among the evaluated models. The SVC followed with 0.0400 g $CO_2$eq, which aligns with its relatively long training duration (9.31 seconds). The Isolation Forest, designed for unsupervised anomaly detection, consumed far less energy ($1.34 \times 10^{-9}$ kWh) and emitted 0.0049 g $CO_2$eq, reflecting its smaller computational footprint despite being tree-based.

In contrast, Logistic Regression and XGBoost were notably more eco-efficient. Logistic Regression emitted only 0.0001 g $CO_2$eq, maintaining near-zero energy usage ($2.73 \times 10^{-11}$ kWh), while XGBoost achieved an excellent balance between accuracy (0.7457) and efficiency, emitting just 0.0025 g $CO_2$eq with minimal computational demand. These results reveal a consistent trend: model complexity directly correlates with energy consumption and emissions. Ensemble-based models such as Random Forest and SVC deliver higher predictive power but at a greater environmental cost. Meanwhile, simpler models like Logistic Regression and optimized gradient-boosting techniques such as XGBoost offer competitive performance with significantly reduced carbon footprints. Although the absolute energy consumption values are small due to the controlled experimental environment, the relative patterns remain meaningful. In real-world cybersecurity operations, where detection systems process continuous network data, these efficiency gaps would scale substantially. Thus, integrating sustainability monitoring tools like CodeCarbon becomes essential for quantifying, comparing, and optimizing the eco-efficiency of cybersecurity AI models.

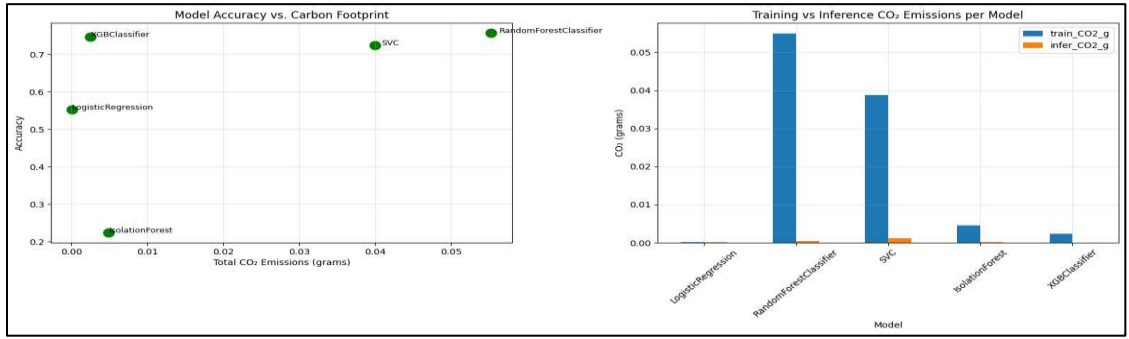

Fig.7: Results of energy consumption and carbon emissions per model

### 4.3 Eco-Efficiency Trade-Offs

The eco-efficiency evaluation combined the F1-score and energy consumption results into a unified index, the Eco-Efficiency Index (EEI), representing the detection quality achieved per kilowatt-hour consumed. While the numerical energy consumption was near-zero in the constrained Colab environment, the relative EcoIndex values effectively illustrated performance-to-energy trade-offs. The Logistic Regression and XGBoost models emerged as the most eco-efficient, achieving strong detection quality relative to their low energy usage. The Random Forest and SVC models, though high-performing, showed lower eco-efficiency due to longer training times and higher emissions. The Isolation Forest's high EcoIndex value was primarily an artifact of minimal energy use rather than detection capability, underscoring that sustainability metrics must always be interpreted alongside performance metrics. Visualization of accuracy versus total $CO_2$ emissions and F1-score versus energy use further revealed the Pareto frontier between performance and sustainability. XGBoost achieved an ideal balance point, combining strong predictive capability with relatively low carbon emissions, suggesting its potential for scalable, real-world, carbon-aware cybersecurity applications. Logistic Regression, meanwhile, represented a lightweight deployment option for energy-constrained or edge computing environments, aligning with sustainable AI principles for low-power systems. [16].

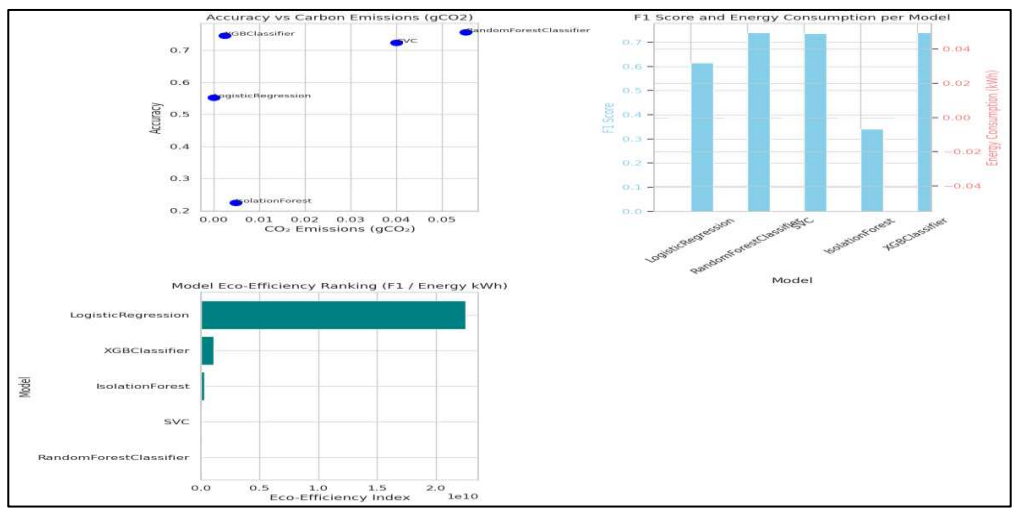

Fig.8: Eco-efficiency evaluation results

### 4.4 Optimization Outcomes

The optimization phase sought to refine model performance while reducing environmental cost through hyperparameter tuning and dimensionality reduction. Optimization led to mixed but revealing outcomes. The Random Forest (optimized) model demonstrated a marginal increase in eco-efficiency despite longer training times, while XGBoost (optimized) preserved its strong detection performance with modest emissions growth. However, the most striking result came from the Random Forest model trained on PCA-reduced features, which achieved the highest overall accuracy (0.7696) and F1-score (0.7532) while reducing training time and $CO_2$ emissions by over an order of magnitude compared to its baseline counterpart. This result validates the hypothesis that feature reduction can serve as an effective sustainability lever, reducing computational effort without compromising model fidelity. PCA condensed redundant or weakly correlated features into a smaller set of principal components, simplifying model complexity and improving both runtime efficiency and environmental footprint. Overall, these results demonstrate that eco-aware optimization strategies, particularly dimensionality reduction, can significantly enhance both detection reliability and carbon efficiency. The findings support the broader principle that sustainable machine learning does not necessitate performance sacrifices; instead, thoughtful design and optimization can yield models that are simultaneously accurate, efficient, and environmentally responsible.

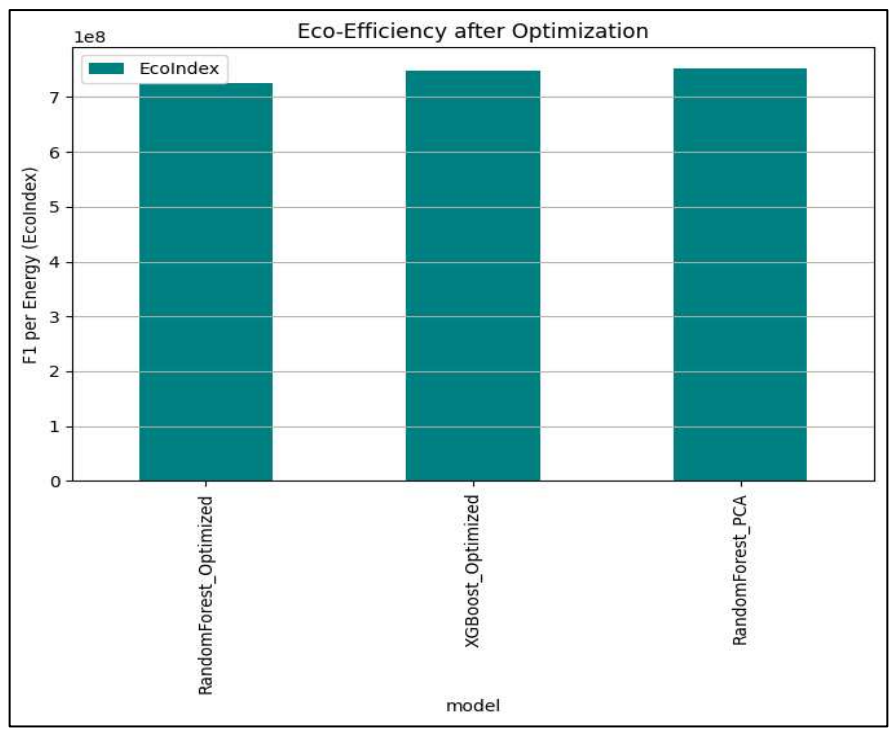

Fig.9: Eco-efficiency results after optimization

### 5. Insights and Implications

Bringing sustainability into cybersecurity marks a turning point for both engineering and environmental practice in the U.S.. As artificial intelligence becomes central to how digital systems operate, its energy use and carbon footprint can no longer be treated as side issues. This study adds to that growing discussion by showing that effective anomaly detection can work hand in hand with lower energy use and reduced emissions, setting the stage for more sustainable cybersecurity systems.

### 5.1 Engineering Significance

From an engineering point of view, this work shows that sustainability metrics can fit naturally into machine learning–based cybersecurity systems without sacrificing detection accuracy. Traditional approaches often focus only on how fast or accurate a model is. Here, energy use and carbon output are treated as equally important measures of performance. By tracking the environmental cost of computation during both training and inference, cybersecurity professionals can start evaluating models not only for predictive strength but also for operational and ecological efficiency. Integrating CodeCarbon into the training workflow proves that sustainable analytics can be a regular part of engineering practice. It gives developers the means to choose models that perform well while consuming less power, effectively introducing a form of carbon-aware decision-making. This approach could lead to adaptive security systems that select models based on both the current threat level and available energy resources. The introduction of the Eco-Efficiency Index (EEI) adds a useful way to compare how efficiently models turn energy into meaningful security outcomes. The results showed that Random Forest and XGBoost maintained high accuracy with solid carbon efficiency, while reduced-feature models revealed opportunities to cut energy use even further. These findings reflect a broader engineering goal: designing defenses that protect data while conserving power. Olivetti et al. (2025) argue that as AI continues to expand across industries, engineers have a responsibility to build systems that are aware of their carbon footprint [19]. This study answers that call, showing that machine learning tools can defend networks while minimizing their own environmental cost.

### 5.2 Environmental Impact

The environmental results offer clear evidence that machine learning models can lower their carbon footprint without losing effectiveness. Even in a controlled setup, differences in energy use across models were striking. Lightweight algorithms such as Logistic Regression and optimized Random Forests consumed much less power than deeper or more complex models. When scaled up to enterprise environments or data centers, where many models run continuously, these differences can translate into substantial energy savings. The findings suggest that AI-driven cybersecurity can help meet environmental goals rather than stand in their way. By including carbon metrics in model evaluation, organizations can align cybersecurity operations with sustainability targets or emission standards. Olivetti et al. (2025) note that the environmental toll of AI comes not only from large training runs but also from ongoing inference and data transfers [19]. Since anomaly detection systems often run around the clock, optimizing their energy use becomes essential. These results also point to opportunities for policy-driven progress. Organizations could begin adopting "green AI" mandates similar to those already present in other high-energy sectors. Requiring environmental reporting in cybersecurity frameworks would create a feedback loop between digital resilience and ecological accountability. In this sense, energy-aware cybersecurity isn't a trade-off but a path toward systems that are both secure and sustainable.

### 5.3 Practical Applications

The practical uses of this research reach across data centers, enterprise monitoring systems, and AI-powered intrusion detection tools. The combination of anomaly detection and energy tracking offers organizations a framework for making smarter, sustainability-informed decisions. In data centers, where energy and security demands are high, carbon-aware machine learning can guide how resources are allocated. During times of high energy demand or low renewable availability, systems could automatically switch to lighter models that still maintain protection while cutting emissions. Network appliances could also log the energy cost of intrusion events, helping security teams visualize how much power their defenses actually consume. The Eco-Efficiency Index can also support compliance and sustainability audits. As U.S. and global initiatives continue to promote low-carbon digital infrastructure, cybersecurity teams could report EEI metrics alongside traditional IT performance indicators. Combining environmental data with security operations supports a more holistic approach to digital governance, where energy accountability becomes part of risk management itself. In academic and training contexts, these findings can inform courses and programs that prepare future cybersecurity professionals to think about both technical rigor and environmental responsibility. The convergence of AI, cybersecurity, and sustainability is not only realistic but increasingly vital to the evolution of secure, responsible digital systems.

### 5.4 Limitations

While this study offers valuable insights, a few limitations frame its scope. The dataset used, though multi-dimensional and realistic, is smaller than what would typically be seen in real-world traffic logs. Testing these methods on larger, more diverse datasets would strengthen the generalizability of the findings. CodeCarbon's emission estimates are based on average regional and hardware factors, meaning they do not fully capture variations in local energy mixes, such as differences between renewable-heavy and fossil-fuel-dependent regions. Future research should incorporate more detailed, location-specific calibration to improve accuracy. The experiments were also limited to a static Colab environment, which cannot reflect the complexities of distributed computing, cooling requirements, or the power interactions seen in enterprise-scale systems. Real-world tests may reveal nonlinear patterns between workload intensity and actual power use. Even with these constraints, the study provides an important starting point for embedding sustainability into cybersecurity research. It lays the foundation for future work that integrates energy-aware learning into large, adaptive systems capable of balancing digital protection with environmental preservation.

### 6. Future Work

As machine learning continues to evolve as a cornerstone of modern cybersecurity, the integration of sustainability metrics opens promising pathways for innovation. The present study demonstrated that anomaly detection systems can be both effective and energy-conscious, but several critical extensions remain to enhance scalability, environmental realism, and operational integration. The next phase of this research will therefore focus on advancing the eco-friendly cybersecurity paradigm toward greater depth, breadth, and practical deployment across heterogeneous computing environments. One essential direction involves integrating renewable energy weighting into the carbon estimation framework. The current approach assumes average U.S. grid emissions, but regional variations in renewable energy penetration, such as those in states with high solar or wind utilization, can significantly alter the carbon cost of computation. Incorporating dynamic carbon intensity factors tied to grid mix data from real-time energy APIs would allow the anomaly detection system to dynamically adjust its sustainability metrics. For example, model selection could shift toward more computationally intensive algorithms when renewable power is abundant, reverting to lightweight models during fossil-heavy hours. This adaptive energy alignment would establish a more authentic link between cybersecurity operations and environmental responsibility, extending the model's utility from mere measurement to real-time carbon optimization.

Another key extension involves expanding the modeling framework to include deep learning and reinforcement learning–based intrusion detection systems (IDS). While the current work focused on classical and ensemble learning models, modern deep architectures such as autoencoders, graph neural networks, and LSTM-based anomaly detectors could capture more complex temporal and relational dependencies in network data. Evaluating these architectures under carbon-aware conditions will clarify whether performance gains justify their environmental cost. Similarly, reinforcement learning could introduce dynamic defense strategies, where the agent balances detection accuracy against energy expenditure, a multi-objective optimization aligned with sustainable computing principles. Equally important is the practical deployment of the framework on low-power edge and IoT devices, which represent the frontier of both cybersecurity and energy management. Testing the system on ARM-based microcontrollers, Raspberry Pi units, or edge gateways will reveal the feasibility of running carbon-aware anomaly detection at the periphery of networks where energy constraints are tightest. This step will bring eco-friendly cybersecurity closer to real-world implementation, especially in smart infrastructure, healthcare IoT, and autonomous systems that demand continuous protection with minimal energy draw.

The future research roadmap also includes multi-objective optimization that explicitly balances accuracy, latency, and carbon cost. Using algorithms such as Pareto optimization or genetic search, it will be possible to identify optimal trade-offs where marginal performance increases no longer justify disproportionate energy use. This aligns with the philosophy of "Green AI," where efficiency becomes as valuable as accuracy. Moreover, the development of open benchmarks and standardized metrics, including the Eco-Efficiency Index proposed in this study, will facilitate consistent comparison across institutions and datasets, fostering a global research community around green cybersecurity AI. Finally, future work will consider the life cycle perspective of artificial intelligence applications. As Plociennik and Lamnatou (2025) emphasize, sustainability in AI must encompass not only runtime energy usage but also the upstream and downstream impacts of data storage, hardware manufacturing, and system disposal [20]. Adopting a life cycle assessment (LCA) approach would enable researchers to account for the full environmental cost of cybersecurity systems, from data collection and model training to deployment and eventual decommissioning. This holistic perspective will help position eco-aware cybersecurity as part of a larger global effort to make AI development transparent, accountable, and environmentally restorative. In conclusion, future work will expand this study from a computational experiment into a comprehensive sustainability framework for cybersecurity. By merging machine learning optimization, renewable energy integration, and life cycle thinking, the next generation of eco-friendly intrusion detection systems will not only secure networks but also align digital security with planetary stewardship.

## Conclusion

This study moves the conversation on cybersecurity in a new direction by bringing environmental awareness into the design of machine learning–based anomaly detection. Using the Carbon-Aware Cybersecurity Traffic Dataset, it connects two fields that rarely overlap, cyber defense and carbon accounting, and shows that choosing a model should now depend not only on how accurate it is but also on how efficiently it uses energy. The introduction of the Eco-Efficiency Index (EEI) offers a practical way to measure the balance between detection performance and energy consumption, giving researchers and engineers a clearer way to think about algorithmic sustainability. The results show that while more complex ensemble models like Random Forest and XGBoost tend to perform better in terms of accuracy, simpler models such as Logistic Regression can achieve a strong balance when energy use is taken into account. Tracking emissions and energy through CodeCarbon gave measurable insight into the real cost of computation, revealing that feature optimization, especially PCA-based dimensionality reduction, can improve both accuracy and energy performance. These findings suggest that security and sustainability can strengthen each other when systems are designed with efficiency in mind.

On a larger scale, this research adds to the growing discussion around "green AI" by making carbon tracking part of cybersecurity analysis, an area that has often been left out of sustainability debates. The work shows that eco-aware security models can realistically be used in data centers and cloud systems, particularly in the U.S., where both policy and industry trends are starting to reward carbon-conscious operations. As cyber threats become more complex and AI workloads continue to expand, scalable detection systems that take their environmental footprint into account will be central to the future of responsible digital security. In essence, this study lays down a framework for thinking about cybersecurity as both a technical and environmental discipline. The Eco-Efficiency Index offers a foundation for future research that blends accuracy, adaptability, and sustainability. By treating energy metrics as part of performance evaluation, the study redefines what success looks like in cybersecurity, pointing toward a future where protecting digital systems also means protecting the planet.